\documentclass[aps,amssymb,showpacs,floatfix, twocolumn, longbibliography]{revtex4-1}
\usepackage{amssymb,graphicx}
\usepackage{epsfig,graphicx}
\usepackage{subcaption}
\usepackage{subfloat}
\usepackage{epstopdf}
\usepackage{amsmath}
\usepackage{xcolor}
\usepackage{romannum}

\begin{document}
\pagenumbering{arabic}

\title{Antiferromagnetism and spin density waves in 3D Dirac semimetals}

\author{Grigory Bednik}

\affiliation{Physics Department, University of California Santa Cruz, CA, USA}

\date{June 13, 2018}

\begin{center}
\begin{abstract}
\newcounter{TypeOne}
\newcounter{TypeTwo}
\newcounter{TypeThree}

\setcounter{TypeOne}{1}
\setcounter{TypeTwo}{2}
\setcounter{TypeThree}{3}

We perform mean-field study of possible magnetic instabilities in Dirac semimetals. We find that Dirac electrons naturally host  antiferromagnetic or spin density wave ground states, though their specific configurations may vary depending on specific model, as well as chemical potential and temperature. We also discuss paramagnetic susceptibility of Dirac semimetals. In the cases, when Dirac electrons do not have orbital momentum, the magnetic properties may be $\mu$ and $T$ independent.


 



\end{abstract}
\end{center}

\maketitle

\section{Introduction}

Weyl and Dirac semimetals (see \cite{RevModPhys.90.015001} for review) have attracted significant attention during the last years. Both of them are prominent for their band structure having effectively gapless excitations described by Weyl/Dirac equation. In the former case, their spectrum contains two non-degenerate bands, and in the latter case their two bands are doubly degenerate. Because of that,  Weyl semimetals are realized, when either time-reversal or inversion symmetry is broken, whereas Dirac semimetals are realized in the presence of both of them.

It is known, that Weyl points in Weyl semimetals \cite{Xu613, PhysRevB.83.205101, PhysRevLett.107.127205, NatCommun6.8373, NatPhys10.3425, Xue1501092, Xu2015} can be viewed as monopoles of Berry curvature \cite{Volovik1987}, which makes them topologically stable. Weyl points cannot be gapped under any local perturbations. The only way to destroy them is via annihilation, which can happen after moving two Weyl points of opposite charges together. In contrast, in Dirac semimetals, the Dirac points are not protected topologically. In type \Roman{TypeOne} Dirac semimetals \cite{PhysRevLett.108.140405, PhysRevB.85.195320, Science343.864, PhysRevLett.113.027603, NatCommun5.4786, NatMatt13.677} (two experimentally discovered examples are $\mathrm{Na_3 Bi}$ and $\mathrm{Cd_3 As_2}$), there are two Dirac points located opposite to each other on the $z$ axis, which are protected by crystal rotational symmetry relative to $z$ axis. At the same time, in type \Roman{TypeTwo} Dirac semimetal $\mathrm{Zr Te_5}$ \cite{PhysRevLett.115.176404, Pariari2017} there is a single Dirac point at the center of the Brillouin zone, which is not protected at all: in fact the Dirac point is gapped, but the gap can be neglected because, accidentally, it is very small.

 Dirac semimetals necessarily possess time-reveral and inversion symmetries, which ensure double degeneracy of their bands.  Breaking one of these symmetries may lift the degeneracy thus splitting the Dirac point  into a pair of Weyl points. It can also lead to more complicated combinations of Weyl points and rings \cite{PhysRevLett.107.127205, PhysRevB.95.161306}. The most natural way to achieve it, is to add intrinsic Zeeman field into the system. The behavior of Dirac semimetals under extrinsic magnetic fields applied in specific directions has been extensively studied, and for instance, it was found that, in the simplest model of $\mathrm{ZrTe_5}$, the Dirac point can split into two Weyl points, if the Zeeman field is applied in $z$ direction, or into a nodal ring if the Zeeman field is applied in $xy$ plane \cite{PhysRevLett.120.016603}.

However, it is not yet fully understood whether magnetization in Dirac semimetals may happen spontaneously. Recently, magnetic properties of Dirac electrons interacting with magnetic impurities have been studied  \cite{PSSB:PSSB201552457, PhysRevB.93.094438, PhysRevB.92.224435, PhysRevB.92.241103}. Magnetic impurities are known to obey  
 RKKY interaction, which has complicated oscillating and anisotropic structure, and thus does not make it possible to find the resulting magnetic ground state. In the works \cite{PhysRevB.93.045201, 1801.10397, PhysRevB.94.195123}, magnetic susceptibility of Dirac semimetals at small external magnetic fields was studied, and a few remarkable properties were found. In particular, it was found that the susceptibility in Dirac semimetals is determined not only by the Fermi surface, but by the whole Brillouin zone, and moreover it may be independent on the Fermi energy, but dependent on the boundary properties of the Brillouin zone. 

 Magnetic instabilities have also been studied in various systems, similar to Dirac semimetals, such as 3D Fermi gas with Weyl-like spin orbit coupling \cite{PhysRevB.94.115121}, Dirac electrons on a surface of topological insulator \cite{PhysRevLett.106.136802, PhysRevB.94.235421, PhysRevB.96.024413, PhysRevB.85.195119, PhysRevLett.115.036805, PhysRevB.89.115101} and particularly Weyl semimetals \cite{PhysRevLett.115.076802, PhysRevB.94.241102, PhysRevB.96.201101, PhysRevB.90.035126, PhysRevB.87.161107, PhysRevB.94.075115}. It was found, that in these systems, various phases can emerge, including spin density waves \cite{PhysRevB.94.115121, PhysRevB.94.235421, PhysRevLett.115.076802}.

On the other hand, in the recent years a large number of antiferromagnetic Dirac semimetals has been discovered experimentally, which include $\mathrm{CuMnAs}$ \cite{PhysRevB.96.224405, Tang2016, PhysRevB.97.125109, PhysRevLett.118.106402}, and also $\mathrm{Ca Mn Bi_2}$, $\mathrm{Sr Mn Bi_2}$ \cite{guo2014coupling, Zhang2016}, and $\mathrm{NdSb}$ \cite{0953-8984-28-23-23LT02, PhysRevB.32.7367, PhysRevB.97.115133, PhysRevB.93.205152}, and $\mathrm{EuCd_2 As_2}$ \cite{PhysRevB.99.245147}. These materials are currently widely explored in the context of possible spintronics applications \cite{Wadley587, PhysRevLett.113.157201, PhysRevLett.92.247201, PhysRevLett.118.106402} (see \cite{doi:10.1002/pssr.201700044} for review). However, their electronic structure is rather complicated and inculdes both localized and conduction electrons. Their phase diagrams are also non-trivial. 

Motivated by this knowledge, we study a simple problem of spontaneous magnetization in 3D Dirac semimetals (which can arise either due to magnetic impurities or due to interactions) using mean field approximation 
\cite{altland2010condensed, fradkin2013field}. Similar approach was previously used to find prefered magnetic states in semiconductors with quadratic band touching \cite{PhysRevLett.84.5628, 10.2307/3074210}. We find that 3D Dirac semimetals may host antiferromagnetic or spin density wave ground states, depending on their Fermi level. 
We observe transitions between different magnetic ground states as functions of Fermi energy and temperature. 

 Specifically, we consider two most commonly used models of 3D Dirac semimetals: a model describing type \Roman{TypeTwo} Dirac semimetal $\mathrm{Zr Te_5}$ with one Dirac point at the center of the Brillouin zone, and a model describing type \Roman{TypeOne} Dirac semimetal $\mathrm{Na_3 Bi}$ with two Dirac points separated in momentum space. We introduce magnetization fields in these models (which may have finite momentum corresponding to spatial modulation of the magnetization), compute their effective actions up to quadratic order and thus demonstrate that the magnetic instabilites are spatially modulated. After it, we place these models on a lattice and compute the effective actions for magnetization numerically. In this way, we are able to see, at which momentum, the effective actions reach minimum, i.e. when magnetizations become stable. We obtain that at small chemical potentials, the minimum of the effective action is reached at the boundary of the Brillouin zone, which implies that the magnetic ground states are antiferromagnetic. On the other hand, as the chemical potential increases, the minimum of the effective action shifts away from the boundary of the Brillouin zone, and thus the magnetic ground state changes into an incommensurate spin density wave. Eventually, there may happen a transition into a ferromagnetic ground state. 

In addition, we compute the effective actions at finite temperature. We obtain their analytical expression for the case of, zero external momentum, which provides us knowledge about paramagnetic susceptibility of the Dirac semimetals. We find that, typically, the paramagnetic susceptibility, decreases $\propto T^2$. However, in certain cases, particularly when Dirac electrons do not have orbital momentum, the susceptibility may be $\mu$ and $T$-independent. Finally, we compute the effective action numerically at finite temperatures and see that the wavevector of the spin density wave changes as a function of temperature. This leads us to conclusion that there may exist transition between commensurate antiferromagnetic and incommensurate spin density wave ground states as a function of temperature.

This paper is organized as follows: in Sec. \ref{Spontaneous_magnetization_at_zero_temperature}, we compute the effective actions for both our models analytically at zero temperature. In particular, in Sec.  \ref{sec:TypeTwoDM} we focus on the case of Dirac semimetal with one Dirac point (i.e. type \Roman{TypeTwo}) and in Sec. \ref{sec:TypeOneDM}, we consider Dirac semimetal with two spatially separated Dirac points (i.e. type \Roman{TypeOne}). In Sec. \ref{Magnetic_susceptibility_at_finite_temperatures}, we compute paramagnetic susceptibility at finite temperatures for both type \Romannum{2} and \Romannum{1} Dirac semimetals. We summarize our findings and their possible implications in Sec. \ref{Sec:Conclusions}. Finally we present our numerical results in the appendix \ref{Sec:Numerical}.


\section{Spontaneous magnetization at zero temperature}
\label{Spontaneous_magnetization_at_zero_temperature}

In this section we study spontaneous magnetization in two simple models of type \Roman{TypeTwo} and \Roman{TypeOne} Dirac semimetals.  
 Specifically, we introduce the magnetization field, compute its effective action up to the second order over the field and its momentum, and use it to find  possible magnetic instabilities.  We start from the simplest model of type \Romannum{2} Dirac semimetal containing a single Dirac point and use it to illustrate our method. After it, we repeat the calculations in the case of slightly more complicated model of type \Romannum{1} Dirac semimetal, which contains a pair of Dirac points separated in momentum space.

\subsection{Type \Roman{TypeTwo} Dirac semimetal}
\label{sec:TypeTwoDM}

Here we analyze the model of type \Roman{TypeTwo} Dirac semimetal, which has a single Dirac point at the center of the Brillouin zone. We note that an experimental example of such Dirac semimetal is $\mathrm{Zr Te_5}$.  Its band structure is invariant under time-reversal, inversion, and also under three mirror symmetries. The Hamiltonian \cite{PhysRevLett.115.176404} can be written as:

\begin{eqnarray}
H_0 =\sum_{k, i} d_i(k) \Gamma_i,
\label{TypeTwoHam}
\end{eqnarray}
where $\Gamma_i$ are five gamma-matrices defined as tensor products of Pauli matrices $\sigma$ and $\tau$ acting in spin and pseudospin spaces, namely 
\begin{eqnarray}
\Gamma_1 = - \tau_z \sigma_z,
\qquad
\Gamma_2= \tau_y,
\qquad
\Gamma_3 =  \tau_z \sigma_x,
\nonumber\\
\Gamma_4 = -\tau_z \sigma_y,
\qquad
\Gamma_5 = -\tau_x.
\nonumber
\end{eqnarray}
The coefficients $d_i$, have, in general, complicated form determined by the whole structure of the Brillouin zone, but near the Dirac point, they can be approximated as linear functions of momentum:
\begin{eqnarray}
d_1 = 0,
\qquad
d_2 = v_z k_z,
\qquad
d_3= v_y k_y,
\nonumber\\
d_4 = v_x k_x,
\qquad
d_5 = m \approx 0.
\label{d_eqs}
\end{eqnarray}
Here $v_{x,y,z}$ are Fermi velocities in all possible directions, which, in general, may be different, since the crystal structure is anisotropic, and $m$ is a gap in the Dirac point, which, strictly speaking exists, but can be neglected  because it is very small. 

In a conventional way (see e.g. \cite{altland2010condensed}), we introduce magnetization by adding to the Hamiltonian an interaction term:
\begin{eqnarray}
H_1 = U \int dx \left( \psi^+ S \psi \right)^2.
\label{H_1}
\end{eqnarray}
We emphasize that since electrons in $\mathrm{ZrTe_5}$ have spin, but do not have an angular momentum, $S$ is just an electron spin, which has a simple expression $S_i = \frac{\sigma_i}{2}$.

After performing Hubbard-Stratonovich transformation, it is possible to get rid of the quartic term (\ref{H_1}) by introducing magnetization field $M(q)$, which is just a superposition of spin matrices: $M(q) = \sum_i b_i(q) S_i$. Thus the total Hamiltonian of our system takes the form:
\begin{eqnarray}
H = H_0 + \frac{U}{4} \sum_q M^2 - \frac{U}{2} \sum_{k,q} \psi^+(k+q) M(q) \psi(k),
\nonumber\\
\label{TotalHam}
\end{eqnarray}
which is just a sum of a free-fermion Hamiltonian, kinetic term for the magnetization field and Zeeman interaction between them. We note that the magnetization field is assumed to have a finite momentum $q$, which corresponds to its spatial modulation. 

The effective action for the magnetization field is obtained by writing partition function for a system with the total Hamiltonian (\ref{TotalHam}) as a path integral and integrating the fermions out. The leading correction to the magnetization kinetic term $M^2$ can be written in terms of the fermionic Green function $G(w, k) = (iw - H_0(k))^{-1}$ as
\begin{eqnarray}
  S = 
\frac{1}{\beta} \sum_{w, k, q } \mathrm{tr}  G(w, k)M(-q)G(w, k+q)M(q) .
\label{GMGMdef}
\end{eqnarray}
Here, $k$ is an internal momentum, and $q$ is the external momentum, corresponding to the spatial modulation of the magnetization. We note that, in this section, we are primarily interested in a possible spontaneous magnetization, but, strictly speaking, the same Eq. (\ref{GMGMdef}) describes momentum-dependent spin susceptibility $\chi_{ij}(q) = - \frac{\partial^2 \delta S_2}{\partial b_i \partial b_j}$.

After using the explicit expressions for $M$ and $G$, performing Matsubara summation and taking the limit of zero external Matsubara frequency, the Eq. (\ref{GMGMdef}) takes the form

\begin{eqnarray}
S
=
\sum\limits_k 
\left\{
\frac{\mathrm{tr} M^2 }{2 (d_+^2 - d_-^2)}
\left[  
d_+ n_F(d_+-\mu) - d_+ n_F(-d_+-\mu) 
\right.
\right.
\nonumber\\
\left.
- d_- n_F(d_--\mu) + d_- n_F(-d_--\mu)
\right]
\nonumber\\
+  
\frac{\mathrm{tr}(d_+\Gamma) M (d_-\Gamma) M }{2(d_+^2 - d_-^2) }
\left[
\frac{n_F(d_+-\mu)}{d_+} - \frac{n_F(-d_+-\mu)}{d_+} 
\right.
\nonumber\\
\left.\left.
- \frac{n_F(d_--\mu)}{d_-} + \frac{n_F(-d_--\mu)}{d_-}
\right]
\right\},
\nonumber\\
\label{TrGMGM_gen_expr}
\end{eqnarray}
where for shortness we have introduced  $d_{\pm} = d(k \pm \frac{q}{2})$.

We proceed further by expanding Eq. (\ref{TrGMGM_gen_expr}) in powers of the external momentum $q$. More specifically, we have to expand $d_{\pm}$ and $n_F(d_{\pm}-\mu), n_F(-d_{\pm}-\mu)$ in powers of $q$. As a result, the effective action (\ref{TrGMGM_gen_expr}) becomes rewritten as a sum of the contributions proportional to $n_F(\pm d -\mu)$ and its derivatives
$\frac{\partial^m n_F(\pm d-\mu)}{\partial^m d}$ respectively. The former contribution can be viewed as a momentum $k$ summation over all filled states and thus labeled as interband, whereas the latter can be viewed as a summation over states at the Fermi surface and thus labeled as intraband.  We note that the fact, that magnetization contains the contribution from outside of the Fermi surface is a special property of Dirac semimetals. We explain this peculiarity by the fact that, in Dirac semimetal, different bands approach very closely each other near the Dirac points, which is in contrast to conventional metals, where a band containing the Fermi surface is assumed to be well-separated from the others. 


Now let us write explicitly the terms, which appear after we expand the effective action (\ref{TrGMGM_gen_expr}) up to the zeroth order over the external momentum $q$. Since at zero temperature, the derivative of Fermi distribution is just a minus delta-function, $\frac{\partial n_F(d - \mu)}{\partial d} = - \delta (d - \mu)$, we can remove the momentum integral in the intraband contribution and thus write the effective action as
\begin{eqnarray}
S^{(0)} =
\frac{1}{2 \pi^2 v_x v_y v_z}  \int dk 
\left\{ - \frac{2 k b_{\perp}^2}{3} - \frac{4 k b_z^2}{3} 
\right\} 
\theta(k - \mu),
\nonumber\\
+
\frac{1}{2\pi^2}
\left\{
- \frac{4 \mu^2 b_{\perp}^2}{3} - \frac{2 \mu^2 b_z^2}{3} 
\right\}
\nonumber\\
\label{Type2_ZeroOrder}
\end{eqnarray}

Interestingly, we find, that the 'bulk' contribution to the effective action is divergent at large $k$. However, this divergence is resolved very easily: we were assuming that the dispersion is linear everywhere and unlimited, whereas in a real material, the range of momenta is limited by the size of the Brillouin zone, and furthermore, the dispersion becomes non-linear far away from the Dirac point. Thus, we can view the integral entering the Eq. (\ref{Type2_ZeroOrder}) as large, but finite, and limited by non-linearities in $d$ and the size of the Brillouin zone. We can also track evolution of the effective action $S^{(0)}$ with change of chemical potential $\mu$. When the Fermi level is aligned with the Dirac points, i.e. at $\mu=0$, the effective action, and consequently the magnetic susceptibility, is solely determined by the bulk of the Brillouin zone. In this special case, the $z$ component of the magnetic field enters with the coefficient of larger magnitude, than $b_{\perp}$, and thus we can infer that magnetic susceptibility in $z$ direction $\chi_{zz}$ is larger than in the perpendicular direction (e.g. $\chi_{xx}$).

However, the contribution to $S^{(0)}$ from the Fermi surface 'competes' against the contribution from the bulk. Indeed, the Fermi surface gives larger contribution to the perpendicular component of the susceptibility, than to its $zz$ component. Thus, at sufficiently large Fermi energies, it is possible that the contribution from the Fermi surface is larger than from the bulk of the Brilluin zone, and as a result, the susceptibility in the perpendicular direction is larger than along $z$ axis, oppositely to the case of small Fermi energy.  

It is of interest to compute the change of effective action over the chemical potential explicitly. In the approximation of linear dispersion, it has the following expression:
\begin{eqnarray}
\Delta S^{(0)}
\mid^{\mu}_{0}
=  -\frac{\mu^2 b_{\perp}^2}{2 \pi^2 v_x v_y v_z}.
\nonumber
\end{eqnarray}
This equation tells us, that the variation of the effective action due to finite chemical potential depends only on perpendicular component of the magnetic field. In other words, in the range  of parameters, where the dispersion can be viewed as linear, the susceptibility in $z$ direction does not depend on the chemical potential. We remark, that independence of observables on chemical potential is a common property of topological semimetals: the fact that 
susceptibility in Dirac semimetals does not depend on chemical potential is similar to the
existence of universal value for anomalous Hall conductivity in Weyl semimetals, which does not depend not only on chemical potential, but even on the presence of superconductivity \cite{PhysRevLett.113.187202, Bednik2016}. 

In order to study possible spatial modulation of the magnetic ground state,
we expand the effective action (\ref{TrGMGM_gen_expr}) up to quadratic order over external momentum $q$. More specifically, after we expand the terms $d_{\pm}$, $n_F(d_{\pm})$ and its derivatives and use 
an identity $\frac{\partial^m n_F(d - \mu)}{\partial^m d} = - \frac{\partial^{m-1}}{\partial^{m-1} d} \delta(d - \mu)$, we obtain the following expression for the quadratic terms in momentum:

\begin{eqnarray}
\nonumber\\
S^{(2)} &=& 
\frac{1}{2 \pi^2 v_x v_y v_z}  \int \frac{dk}{k} 
\left\{
 - \frac{q_z^2 b_{\perp}^2}{6} + \frac{q_{\perp}^2 b_z^2}{3}  
\right.
\nonumber\\
&& \left.
  + \frac{  (q_x b_x + q_y b_y)^2  }{6} 
- \frac{  (q_x b_y - q_y b_x)^2  }{6} 
\right\} 
\theta(k - \mu)
\nonumber\\
\-
&+&
\frac{1}{2 \pi^2 v_x v_y v_z}
\-\-
\left\{
- \frac{q_{\perp}^2 b_{\perp}^2}{180} + \frac{43 q_z^2 b_{\perp}^2 }{180}
- \frac{19q_{\perp}^2 b_z^2 }{90} + \frac{q_z^2 b_z^2}{30}
\right.
\nonumber\\
&&
\left.
\-
- \frac{ 11 (q_x b_x + q_y b_y)^2  }{90} 
+ \frac{ 11 (q_x b_y - q_y b_x)^2  }{90} 
\right\}.
\nonumber\\
\label{S_2^2Expansion}
\end{eqnarray}

 Similarly to  Eq. (\ref{Type2_ZeroOrder}), this equation has two contributions from the bulk of the Brillouin zone and from the Fermi surface respectively. They contain terms entering with the opposite signs, which implies that the bulk and the Fermi surface 'compete' against each other, and as a result, there exists a possibility of crossovers between different phases at different Fermi energies. Specifically, at small Fermi energies, 
the 'bulk' contribution dominates, since the momentum integral in the Eq. (\ref{S_2^2Expansion}) is determined by the UV cutoff of the Brillouin zone. However, this contribution decreases with increasing $\mu$, whereas the Fermi surface contribution does not depend on the Fermi level $\mu$, and thus, at sufficiently large $\mu$, can become  dominant.

Now, let us discuss the form of possible magnetic configurations. First, we consider the magnetic instability in $z$ direction. At small Fermi energies, the magnetic field component $b_z$ enters the effective action with positive coefficient in front of the momentum, which implies that the magnetization in $z$ direction is spatially independent. We note, however, that this is due to the fact that the bulk contribution does not have a term $q_z^2 b_z^2$, which in turn happens due to linear dispersion. In principle, non-linear corrections to $d(k)$ can lead to non-trivial 'bulk' term $q_z^2 b_z^2$, which in turn may change the ground state into spin density wave
 modulated in $z$ direction, as we will see in numerical calculations.
In addition, at sufficiently large Fermi energy, the term $q_{\perp}^2 b_z^2$ may acquire negative coefficient, which means that magnetic field $b_z$ will be spatially modulated in the perpendicular direction. 

In a similar way, if we look at the instability in perpendicular direction $b_{\perp}$,  the Eq. (\ref{S_2^2Expansion}) tells us that, at small Fermi energy, it will be spatially modulated in the transverse direction, since the effective action has negative
momentum components in $z$ direction and in the $xy$ plane perpendicular to the magnetization. If the Fermi level is increased, its momentum may change: $q_z$ component may disappear, and, furthermore, it may change from transverse to longitudinal, i.e. the momentum vector may become aligned with the magnetization direction.

Thus, we have found that the effective action for magnetization in type \Romannum{2} Dirac semimetal does not reach its local minimum at zero external momentum $q$. This fact tells us that the magnetic state of the type \Romannum{2} Dirac semimetal is spatially modulated, and its prefered configuration is determined by the competition between the 'bulk' (i.e. interband) and the Fermi surface (i.e. intraband) contributions to the effective action. As a result, the magnetic ground state may change as the Fermi level changes. Since we are able to compute analytically the effective action only in the limit of small momenta $q$, we cannot obtain an explicit answer for a wavelength of the spatial modulation. For this reason, in the appendix \ref{Sec:Numerical} we compute the effective action (\ref{TrGMGM_gen_expr}) numerically at arbitrary values of the external momentum $q$ and obtain that at small Fermi level, its minimum is reached at the boundary of the Brillouin zone, i.e. when one of the components of $q$ is equal to $\pm \pi$. However, when the Fermi level becomes sufficiently large and non-linear corrections to the Hamiltonian (\ref{TypeTwoHam}) become important, the minimum shifts away from the boundary of the Brillouin zone. Thus, we claim that at small Fermi level, our model of Dirac semimetal has an antiferromagnetic ground state, but as the Fermi level increases, it undergoes phase transition to an incommensurate spin density wave ground state. 



\subsection{Type \Roman{TypeOne} Dirac semimetal}

\label{sec:TypeOneDM}

Type \Roman{TypeOne} Dirac semimetal has two experimentally discovered examples: $\mathrm{Na_3 Bi}$ and $\mathrm{Cd_3 As_2}$. It possesses a pair of Dirac points separated in momentum space in $z$ direction and protected by discrete rotational symmetry. We write the simplest Hamiltonian describing such a system as
\begin{eqnarray}
H_0 = \sum_{k, i} d_i (k) \Gamma_i,
\label{HamiltonianTypeOne}
\end{eqnarray}
and, to be consistent with the previous literature\cite{PhysRevB.95.161306}, we define the $\Gamma$-matrices in terms of Pauli matrices $\sigma, \tau$ as
\begin{eqnarray}
\Gamma_1 =  \tau_z \sigma_x,
\qquad
\Gamma_2= \tau_z \sigma_y,
\qquad
\Gamma_3 = \tau_z \sigma_z,
\nonumber\\
\Gamma_4 = \tau_x,
\qquad
\Gamma_5 = \tau_y.
\nonumber
\end{eqnarray}
We take the simplest possible form of the coefficients $d_i$:
\begin{eqnarray}
d_1 = v_F k_x,
\qquad
d_2 = v_F k_y,
\qquad
d_3 = m(k_z)
\nonumber\\
d_4 = d_5 = 0
\nonumber
\end{eqnarray}
Here $m(k)$ is a function, that  
changes sign at two symmetric points separated in $z$ direction, which are indeed the Dirac points. We assume that $m$ is positive between the Dirac points (e.g. at $k=0$) and negative away from them. 

In type \Romannum{1} Dirac semimetal, the Dirac points are protected by discrete rotational symmetry \cite{NComms5.5898} along the $z$ axis. Namely,  states in the valence and conduction bands have different rotation eigenvalues, which makes it impossible to write a rotationally-invariant term, that would gap them out. This, in turn, results from the fact that, in type \Roman{TypeOne} Dirac semimetals, the valence and conduction band belong to different multiplets: the valence band is a singlet with total spin $J=1/2$, whereas the conduction band is part of the triplet with total spin $J=3/2$ (see e.g. \cite{PhysRevB.85.195320, PhysRevB.95.161306} for more details). The other bands from the triplet are separated by an energy gap, so we neglect them.  Thus we start our analysis from writing a Hamiltonian of the form (\ref{TotalHam}) with $H_0$ describing free electrons in type \Roman{TypeOne} Dirac semimetal (see Eq. \ref{HamiltonianTypeOne}),  but we write the magnetization operator as a sum of singlet and triplet contributions
\begin{eqnarray}
M = M_s + M_p,
\nonumber
\end{eqnarray}
which  
have matrix structure following from the total angular momentum of the states in the bands, namely,
\begin{eqnarray}
M_s = \frac{g_s}{2} 
\left(
	\begin{array}{cccc}
	0  &  0     &  0    &  0  \\
	0  & b_z  & b_-  &  0  \\
	0  & b_+ & -b_z&  0  \\
	0  &  0	    &  0	    &  0
	\end{array}
\nonumber
\right),
\end{eqnarray}
and 
\begin{eqnarray}
M_p = 
\frac{3g_p}{2}
\left(
	\begin{array}{cccc}
	b_z  &  0     &  0    &  0  \\
	0  & 0  & 0   &  0  \\
	0  & 0 &  0&  0  \\
	0  &  0	    &  0	    &   -  b_z
	\end{array}
\right).
\nonumber
\end{eqnarray}
We note that we include in the
 triplet contribution  only $z$ component of the magnetization field. This is because interaction with $b_{x, y}$ can occur only through mixing of the different bands 
within the triplet, which are separated by an energy gap,
but we are interested in scales smaller than the width of each band. We also note, that since, the conduction and valence band belong to different multiplets, they have, generically, different gyromagnetic factors, which we denote by $g_{s,p}$ respectively.

We can compute the effective action (\ref{GMGMdef}) in the same way, as in Sec. \ref{sec:TypeTwoDM}. More specifically, we can still use its expression (\ref{TrGMGM_gen_expr}) and after explicitly evaluating the traces we  obtain that, at zero external momentum $q$, it  can also be represented as a sum of contributions independent and proportional to the Fermi level respectively:

\begin{eqnarray}
 S^{(0)} = - \int \frac{d^3 k}{(2\pi)^3} \frac{v_F^2 k_{\perp}^2}{8E^3}
\left\{ 
(3g_p - g_s)^2 b_z^2  +  g_s^2 b_{\perp}^2
\right\}  
\nonumber\\
- \frac{\mu^2}{8 \pi^2 v_F^2 v_z}
\left\{
	(3 g_p + g_s)^2 b_z^2  +  g_s^2 b_{\perp}^2
\right\}
\nonumber\\
\label{S_2^0_type_1}
\end{eqnarray}
We note, that we have obtained the last row of this equation in the limit of small Fermi level, i.e. when we can view dispersion within each Dirac cone as linear in all directions. 

If we  assume that the magnetic factors $g_{s, p}$ are close to each other: $g_s \approx g_p$, we can conclude that the coefficient in front of $b_z^2$ has larger magnitude than the contribution to $b_{\perp}^2$, and thus magnetic susceptibility will be larger in $z$ direction. Also, we can notice that, as the chemical potential increases, susceptibility always gets  enhanced.  
This result is different from type \Roman{TypeTwo} Dirac semimetal, whose susceptibility in one of directions does not depend on the  the magnitude of the chemical potential. However, it is  worth pointing out a special case $g_s = \pm 3g_p$, which could, in principle, happen if the bands did not have orbital momentum and their pseudospin $\sigma$ or $\tau$ were physical spin. Indeed, in such a case,
 the terms in the Eq. (\ref{S_2^0_type_1}) proportional to $b_z$ become trivial, and thus the magnetic susceptibility becomes independent on the chemical potential.

Now, let us explore spatial modulation of the spontaneous  magnetization. In the same way, as in the Sec. \ref{sec:TypeTwoDM}, we compute quadratic over external momentum $q$ corrections to the effective action and obtain the following expression:

\begin{eqnarray}
  S^{(2)} &=&  - \int\limits_{|m|> \mu} dk_z \frac{g_s g_p  v_F^2 q_{\perp}^2 b_z^2}{4 v_F^2 |m|}
\nonumber\\
&&  - \int \limits_{|m|> \mu} dk_z \frac{m'^2 q_z^2}{48 v_F^2 |m|} 
\left[
	(3 g_p + g_s)^2 b_z^2 +  g_s^2 b_{\perp}^2
\right]
\nonumber\\
&&  + \int \limits_{|m|> \mu} dk_z \frac{m m'' q_z^2}{32 v_F^2 |m|} 
\left[
	(3 g_p - g_s)^2 b_z^2 +  g_s^2 b_{\perp}^2
\right]
\nonumber\\
&&  + \frac{ \pi v_F^2 q_{\perp}^2}{v_F^2 v_z} 
\left[
	\frac{3 g_p^2 b_z^2}{2} + \frac{g_s^2 b^2}{3} - \frac{59 g_s g_p b_z^2}{10}
\right]
\nonumber\\
&&  + \frac{\pi}{v_F^2 v_z} \frac{v_z^2 q_z^2}{12} 
\left[
	(3g_p - g_s)^2 b_z^2 + g_s^2 b_{\perp}^2
\right].
\nonumber
\label{S_2^2_type_1}
\end{eqnarray}

We can see, that similarly to type \Roman{TypeTwo} Dirac semimetal, $S^{(2)}$ contains momentum integral, which diverges in the limit of zero $\mu$, and thus becomes the dominant if $\mu$ is small. However, as $\mu$ grows, its contribution decreases, whereas the other (i.e. intraband) contribution remains invariant. Thus, we can infer, that the $z$ component of the magnetic field is spatially modulated in both $z$ and perpendicular direction, but it can undergo crossovers, as $\mu$ is increased. Similarly, the perpendicular magnetization $b_{\perp}$, if present, is spatially modulated in $z$ direction, and again, with  increasing $\mu$, it can undergo crossover. In the appendix \ref{Sec:Numerical} we present numerical results obtained from lattice regularization of the Hamiltonian (\ref{HamiltonianTypeOne}), which confirm our analytical calculation and also demonstrate that at small $\mu$ the spatially modulated magnetic state is antiferromagnetic, but as $\mu$ increases, it may  undergo transition to a spin density wave state.  

Overall, we have found that Dirac semimetals naturally host spatially modulated magnetic ground states, though their specific configurations may depend on details of the model. Furthermore, we have found that the ground state is determined by the competition between the bulk of the Brillouin zone and the Fermi surface, and therefore can change with Fermi level. In the appendix \ref{Sec:Numerical} we consider the same problem numerically on a lattice and obtain that in most cases, the ground state is antiferromagnetic, though it can also undergo transitions to a spin density wave or a ferromagnet.



\section{Magnetic susceptibility at finite temperatures}
\label{Magnetic_susceptibility_at_finite_temperatures}

In the previous section, we studied possible magnetic instabilities in Dirac semimetals at zero temperature and found that spontaneous magnetization occurs at finite wavevector, thus forming antiferromagnetic or spin density wave phase. Now we would like to get an idea of how it may possibly change, once 
the temperature becomes finite. Since at both finite temperature and wavevector,
the effective action cannot be computed analytically,  
we limit our analysis to the case of zero wavevector.

We start from Eq. (\ref{TrGMGM_gen_expr}). Since we are only interested in the change of the effective action due to small temperature $T << \mu$, we leave only terms, which contain $n_F (d - \mu)$ or its first derivative. After taking the limit $q \to 0$, we obtain:
\begin{eqnarray}
\Delta S_2^{(0)}
=  \int \frac{d^3 k}{(2\pi)^3}
\left\{
 \frac{\mathrm{tr} M^2}{2}
\left(   \frac{n_F(d-\mu) }{2d}   +   \frac{1}{2} \frac{\partial n_F(d-\mu)}{\partial d}  \right)
\right.
\nonumber\\
+
\left.
\frac{\mathrm{tr} (\vec{d}_k \vec{\Gamma}) M (\vec{d}_k \vec{\Gamma}) M}{2}
\left(  -\frac{ n_F(d-\mu)}{2d^3}  +  \frac{1}{2d^2} \frac{\partial n_F(d-\mu)}{\partial d}  \right)
\right\}
\nonumber\\
\label{EffectiveActionFiniteTemperature}
\end{eqnarray}
For each of the models, we can substitute explicit expressions for the traces, and then take the momentum integrals (see e.g. \cite{landau2013statistical} for details of computing integrals over Fermi distributions). However, we also have to account for the fact that $\mu$ is  a function of temperature. This $T$-dependence (at $T << \mu$) can be explicitly found by imposing the condition of conserved particle number $N(T) = \mathrm{const}$ and computing it explicitly:
\begin{eqnarray}
N_{per \: node} = \int \frac{d^3 k}{(2\pi)^3} n_F(k-\mu) 
\approx \frac{\mu^3 + \pi^2 T^2 \mu}{6 \pi^2},
\nonumber
\end{eqnarray}
which in turn leads to the following expansion for the chemical potential:
\begin{eqnarray}
\mu \approx E_F - \frac{\pi^2 T^2}{3 E_F}
\nonumber
\end{eqnarray}
Note, that the expression for $\mu$ does not depend on the number of Weyl nodes. 

 We can evaluate the effective action (\ref{EffectiveActionFiniteTemperature}) by using explicit expressions for the traces, taking momentum integrals and using the above expansion for $\mu$. In the case of type \Roman{TypeTwo} Dirac semimetal, we obtain the following answer:
\begin{eqnarray}
\Delta S_2^{(0)} = 
- \frac{b_{\perp}^2 E_F^2}{2 \pi^2 v_x v_y v_z} 
+
\frac{b_{\perp}^2 T^2}{6 v_x v_y v_z}.
\nonumber
\end{eqnarray}
The most important feature of this expression is that it does not contain any dependency on $b_z$. In other words, we have obtained a surprising result, that in the leading order, paramagnetic susceptibility of type \Romannum{2} Dirac semimetal in $z$ direction is temperature-independent. This is the consequence of the fact that, in Dirac semimetal, bands are very close to each other, which makes interband contribution to the susceptibility (Van-Fleck paramagnetism) comparable to the intraband contribution (Pauli paramagnetism). Moreover,  we were assuming here that the dispersion is linear, which is just an approximation in real materials. However, since we obtained that the susceptibility is temperature-independent, it seems that by slight perturbing, it is possible to make it either decreasing or increasing with temperature.

In type \Romannum{1} Dirac semimetal, the answer for the effective action has the form:
\begin{eqnarray}
\Delta S_{2}^{(0)}
=
- \frac{1}{8 \pi^2 v_{\perp}^2 v_z}
\left\{
	(3 g_p + g_s)^2 b_z^2 + g_s^2 b_{\perp}^2
\right\}
\left(
	E_F^2 - \frac{\pi^2 T^2}{3}
\right).
\nonumber
\end{eqnarray}
As we can see, the susceptibility is temperature-dependent in all directions. However, we note that it becomes temperature- independent in $z$ direction in the special case $g_s = - 3 g_p$, which could, in principle, happen if the band pseudospin were the same as physical spin.

To summarize our results, we have studied magnetic susceptibility of Dirac semimetals as a function of temperature and found that, in certain cases (namely, when pseudospin coincides with physical spin), it may be temperature-independent. In this section, we did not study the behavior of the antiferromagnetic or spin density wave istabilities at finite temperature, but we  consider it numerically in the appendix \ref{Lattice_finite_temperature}. We obtain that the ground state evolves, and, in principle, it may undergo a transition, e.g. from spin density wave to antiferromagnetic phase. 


\section{Discussion}
\label{Sec:Conclusions}

In this paper, we have explored a possibility of spontaneous magnetization in Dirac semimetals, and found that they naturally host antiferromagnetic or spin density wave ground states.  We have also found that their specific structure may vary depending on a particular type of Dirac semimetal, as well as its Fermi energy, temperature, non-linear corrections to the Dirac spectrum. However,  in special cases, when the bands do not have orbital momentum, magnetic properties of Dirac semimetals, including magnetic susceptibility, may be Fermi energy and temperature- independent. 



The main reason, which makes Dirac semimetals different from other solids lies in the fact that, in contrast to conventional Fermi-liquid, magnetization in Dirac semimetal is created not only by electrons on the Fermi surface, but also by electrons from the whole Brillouin zone. In the limit of small chemical potential, i.e. when the Fermi level is close to the Dirac points, the Fermi surface gets reduced to a point(s), and thus the magnetization is created mainly by the bulk of the Brillouin zone. On the other hand, at finite Fermi level, the magnetization arises from competition between the 'bulk' and the Fermi surface contribution, which results in a possibility of phase transitions with varying Fermi energies.

We expect, that our findings may have a lot of implications. 
In fact, the problem of spatially inhomogeneous magnetization in Weyl/Dirac semimetals has also been extensively studied \cite{Tang2014, PhysRevX.6.041046, BednikGrigory2018}, and it was found that it may lead to unusual properties. For example, in the presence of periodic magnetization, there appear novel electronic states, so-called pseudo-Landau levels, which have dispesion forming an 'open nodal line'. We suggest that such effects may arise due to spin density waves in Dirac semimetals, which we study in this work.

 We note that type $\Roman{TypeTwo}$ Dirac cone has been experimentally observed in a material $\mathrm{Zr Te_5}$ \cite{Pariari2017}. In fact, this material has been studied for a long time  \cite{doi:10.1143/JPSJ.49.839}. First, it was discovered as a material, which exhibits anomalous resistivity peak at $T=150K$ \cite{doi:10.1143/JPSJ.49.839}, but soon after, it was claimed that such an anomaly is not attributed to spin or charge density wave \cite{doi:10.1143/JPSJ.51.460}.  We suggest that such a conclusion may be reconsidered, for example, because in the work \cite{doi:10.1143/JPSJ.51.460} it was implicitly assumed that spin/charge density wave  leads to suppresion of carrier densities at Fermi level, but it may not be the case (e.g. if there appear gapless pseudo-Landau levels). 

We mention that, in the recent years, antiferromagnetism was found to be common in Dirac and Weyl semimetals. A large number of materials, where antiferromagnetism coexists with Dirac electrons, was discovered. These materials include, for example, antiferromagnetic NdSb \cite{PhysRevB.93.205152, PhysRevB.97.115133}, where localized spins form ferromagnetic planes, which are antiferromagnetically aligned in one of the directions - similarly to the models considered in this paper. Weyl points were also theoretically predicted in antiferromagnets $\mathrm{Mg_3 Sn}$ and $\mathrm{Mg_3 Ge}$ \cite{PhysRevLett.119.087202, 1367-2630-19-1-015008}. More interestingly, in Ref. \cite{Zhang2016} it was claimed that Dirac electrons enhance antiferromagnetic exchange interaction in the experimentally discovered Dirac semimetals $\mathrm{CaMnBi_2}$ and $\mathrm{SrMnBi_2}$. Another example of Weyl semimetal, which contains ferromagnetic planes aligned antiferromagnetically is $\mathrm{Ba Mn Sb_2}$ \cite{Huang6256}. Perhaps, the most interesting material, where Dirac electrons coexist with antiferromagnetism is $\mathrm{CuMnAs}$ \cite{PhysRevB.96.224405, Tang2016}: in this material, phase transition between commensurate and incommensurate antiferromagnetism has been observed as a function of temperature and chemical composition (i.e. chemical potential) \cite{PhysRevB.96.224405}. Similarly, a material $\mathrm{EuCd_2 As_2}$ \cite{PhysRevB.99.245147} has been predicted to be antiferromagnetic, but it may undergo a transition into a ferromagnetic phase under doping. Finally we mention Dirac material $\mathrm{Sr_{1-y} Mn_{1-z} Sb_2}$   \cite{Liu2017}, which exhibits so-called canted antiferromagnetic order, i.e. two spin components are antiferromagnetically ordered in such a way that the net magnetization is non-zero, but at higher temperatures it undergoes  transition to a ferromagnetic phase. 
Interplay between antiferromagnetism and  Dirac electrons is currently being actively studied, and it leads to novel effects, which have promising applications in spintronics \cite{doi:10.1002/pssr.201700044}. 

Here we have presented a simple mean-field picture, explaining why antiferromagnetism naturally appears in Dirac semimetals. We remark that our approach is not to be viewed as rigorous: mean-field is just a rough approximation, which does not always predict quantities (e.g. temperature dependencies) accurately. A significant contribution to the magnetization behavior may arise due to quantum corrections (e.g. magnons), the full electronic structure etc. In addition, rigorously speaking, magnetic instabilities in Dirac semimetals have to be compared with various instabilities of of different types (see e.g. \cite{PhysRevB.95.201102}). Nevertheless, we have demonstrated that simple mean-field picture successfully explains the origin of antiferromagnetism in Dirac semimetals.

In the future, it might be interesting to derive the same results using more rigorous techniques, e.g. RG analysis. Finally, we note that while preparing this manuscript, we became aware of an experimental work \cite{arXiv:1909.04037}, where charge density waves were found in closely related Weyl semimetals. We believe that they may be described using the same method as spin density waves considered in the present work. 


\begin{acknowledgements}
The author would like to thank Anton Burkov for numerous discussions about this work. The author also would like to thank Sergey Syzranov, Susanne Stemmer, Adam Kaminski, Yun Wu for making useful suggestions and feedback about the project. The work was supported by NSERC of Canada. 
\end{acknowledgements}


\appendix
\section{Numerical calculation of possible magnetic ground states in  Dirac semimetal}
\label{Sec:Numerical}

In this appendix, we consider lattice versions of the models of type \Romannum{2} and \Romannum{1} Dirac semimetals introduced in the main text, 
and compute their effective actions (Eq. \ref{GMGMdef}) numerically. 
In this way, we are able to find the momentum $q$ of the magnetic ground state (determined by a minimum of the effective action $S(q)$) and its behavior at different $\mu$ and $T$, including the range, when non-linear corrections to the dispersion become important. We obtain that, typically, at small $\mu$, the momentum $q$ has one of the components equal to $\pi$, which tells us that the ground state is antiferromagnetic, whereas at sufficiently large $\mu$ (when non-linear corrections become important), $q$ starts decreasing, thus 
showing a phase transition from antiferromagnetic to spin density wave ground state. We note that, throughout our analysis, we do not consider the full 3D range of momenta, which would be computationally challenging, but instead we limit ourselves with just a few special directions of $q$, which is sufficient for our illustrative purposes. 

This section is organized as follows. In the Sec. \ref{Sec_magnetic_gs}
 we describe our method of finding the magnetic ground state at zero temperature by considering the model of  type \Romannum{2} Dirac semimetal, and then repeating 
the calculations for the case of type \Romannum{1} Dirac semimetal. After it, in Sec. \ref{Lattice_finite_temperature} we generalize our method in the case of 
finite temperatures. 


\subsection{Magnetic ground states}
\label{Sec_magnetic_gs}


Let us describe our approach of finding the magnetic ground states by using an example of type \Roman{TypeTwo} Dirac semimetal. We replace the coefficients $d_i$ from the  Eq. (\ref{d_eqs}) with their lattice counterparts, which, in type \Roman{TypeTwo} case, have the form:
\begin{eqnarray}
d_2 = v_z \sin k_z,
\qquad
d_3= v_y \sin k_y,
\nonumber\\
d_4 = v_x \sin k_x.
\label{Type_2_DM_Lattice}
\end{eqnarray}

\begin{figure}
\centering
	\begin{subfigure}[t]{0.5\textwidth}
		\includegraphics[width=9cm,angle=0]{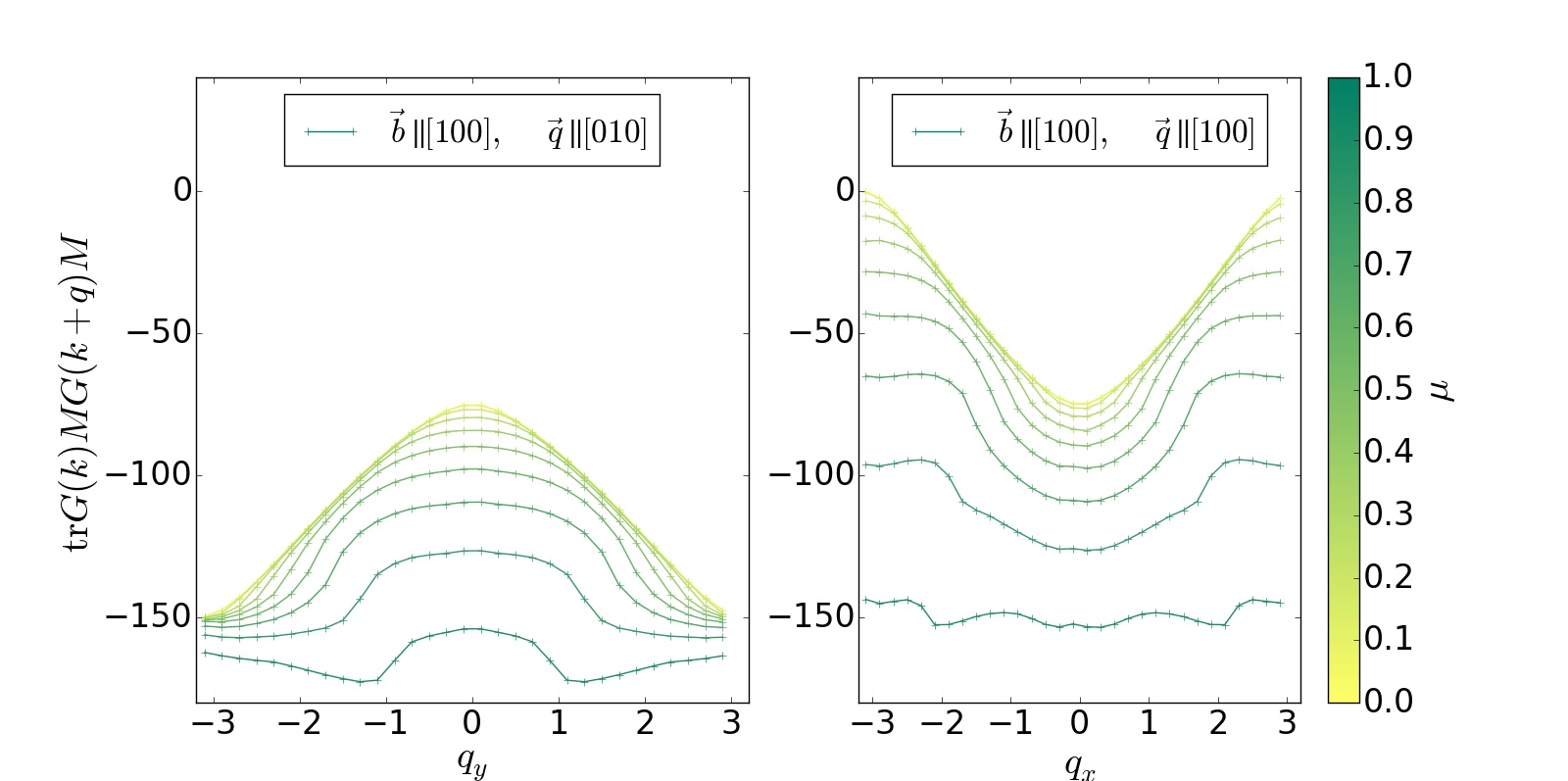}
		\subcaption{$\vec{b} \parallel \vec{x}$}	
		\label{Bx_plot}
	\end{subfigure}
	\begin{subfigure}[t]{0.5\textwidth}
		\includegraphics[width=9cm,angle=0]{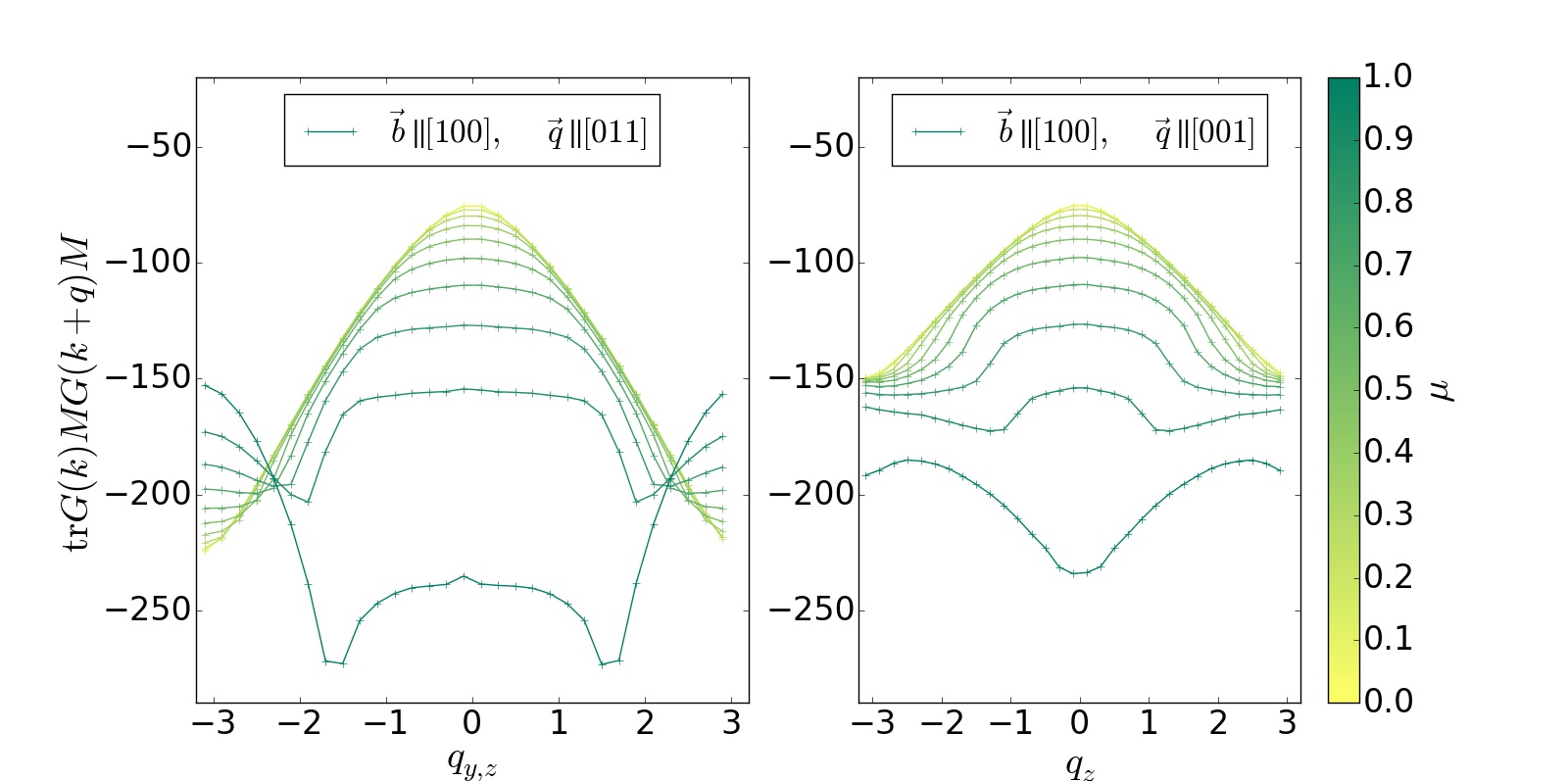}
		\subcaption{$\vec{b} \parallel \vec{x}$}	
		\label{Bx_plot_2}
	\end{subfigure}
	\begin{subfigure}[t]{0.5\textwidth}
		\includegraphics[width=9cm,angle=0]{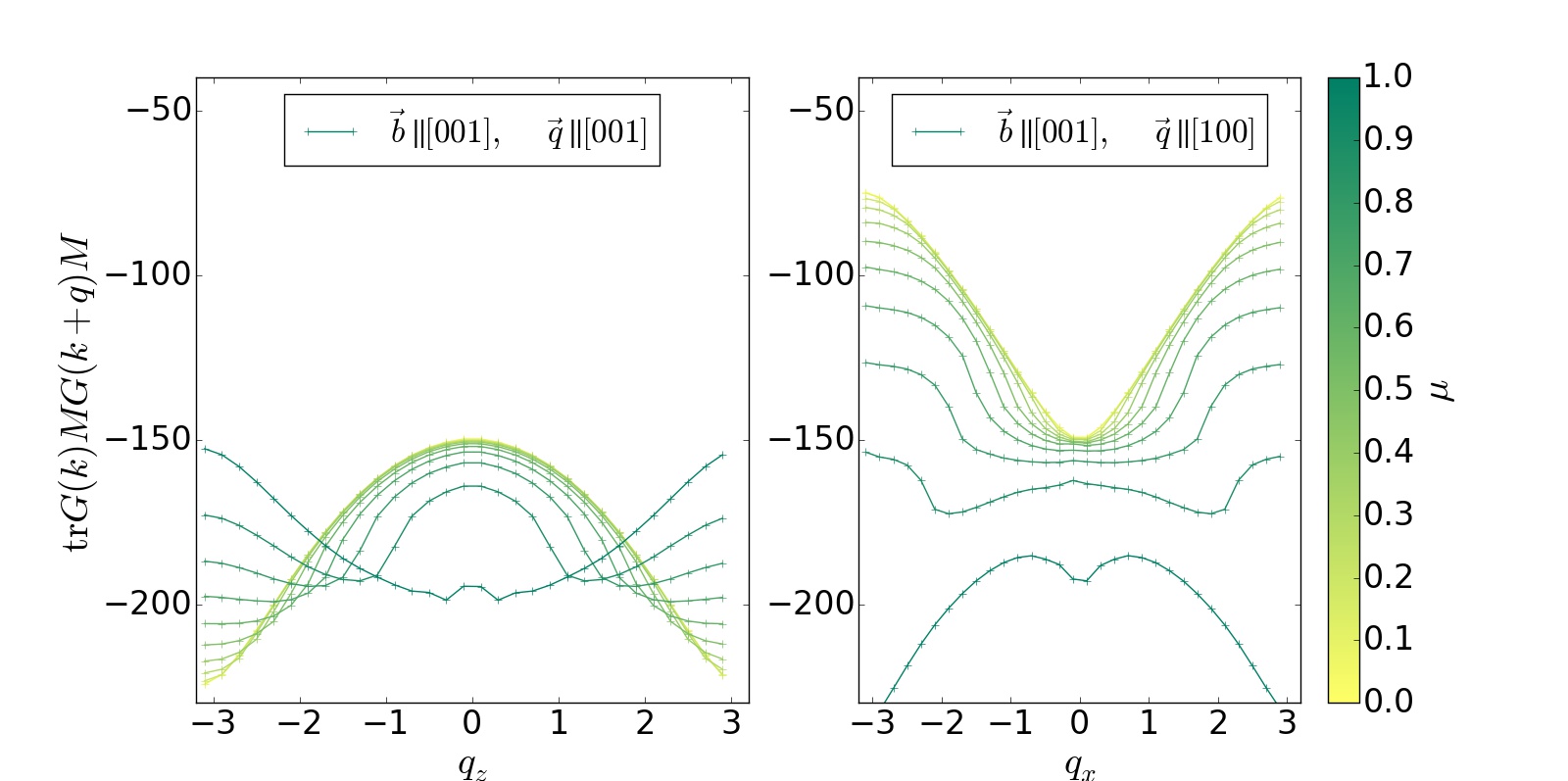}
		\subcaption{$\vec{b} \parallel \vec{z}$}	
		\label{Bz_plot}
	\end{subfigure}
\caption{
	Numerically computed effective action for the model (\ref{Type_2_DM_Lattice}) in the case of: (\ref{Bz_plot})
	$\vec b = (0, 0, 0.5)$ as functions of $\vec{q}$ in the form  $\vec{q} =  (0,0,q_z)$ and  $\vec{q} =  (q_x,0,0)$; (\ref{Bx_plot}) $\vec b = (0.5 , 0, 0)$ as functions of $\vec{q}$ in the form  $\vec{q} =  (0,q_y,0)$ $\vec{q} =  (q_x,0,0)$; (\ref{Bx_plot_2}) $\vec b = (0.5 , 0, 0)$ as functions of $q = q_y=q_z$ and $\vec{q} =  (0,0,q_z)$. The plots are aligned vertically, so that once can see that, in the ground state,  (\ref{Bz_plot}) $b_z$ is modulated in $z$ direction, whereas (\ref{Bx_plot}) $b_x$ is modulated in $y$ direction. Also, in the fig. \ref{Bz_plot} one can see a special case of $\mu = 1.001$, when $b_z$ is modulated in $x$ direction. Thus, we conclude that as $\mu$ reaches band energy at the point $(\pi, 0, 0)$, phase transition occurs. 
}
\label{Fig:TypeTwo}
\end{figure}

 We use these expressions to compute numerically the effective action (see Eq. \ref{TrGMGM_gen_expr}). More specifically, we 
evaluate the momentum integral entering the effective action (\ref{TrGMGM_gen_expr}) over the Brillouin zone (i.e. over the range $-\pi < k_{x,y,z} < \pi$) by using tetrahedral method \cite{PhysRevB.11.2109}. Its main idea is that after the integration range is split into cubes by discretizing momentum, each of the cubes is additionally split into nine tetrahedra. This approach helps us resolve issues, which may happen due to divergences near the Fermi surface in the numerator and denominator of Eq. \ref{TrGMGM_gen_expr}. We present our findings on the Fig. \ref{Fig:TypeTwo}. 
One can see that for some directions of the momentum $q_i$, the effective action has minimum at zero $q_i$,
as was predicted by the Eq. (\ref{S_2^2Expansion}), whereas for others, the minimum is reached at finite values of $q_i$. For example, in the case of $b \parallel x$ (see Fig. \ref{Bx_plot}), the effective action increases with $q_x$, and the curvature decreases with increasing $\mu$ consistently with the Eq. (\ref{S_2^2Expansion}). At the same time, the effective action describing $b_x$ as a function of $q_{y, z}$ (see Fig. \ref{Bx_plot_2}) has minimum away from zero. Thus, $b_x$ component of the magnetization is spatially modulated both in $y$ and $z$ directions.

In a similar way, one can see that magnetization $b_z$ (see Fig. \ref{Bz_plot}) is spatially modulated in $z$ direction (as we will see, this is due to non-linear corrections to the energy spectrum).  At small values of the chemical potential $\mu$, the effective action  has a  minimum at the boundary of the Brillouin zone, which implies that the ground state is antiferromagnetic. However, at larger values of $\mu$, spatial modulation of the ground state decreases, so that it starts forming spin density wave. Eventually, spatial modulation in the $z$ direction disappears, but, on the other hand, it appears in $x$ direction. Thus, as $\mu$ increases, there appears a phase transition between two spin density waves modulated in different directions. 

We note that our numerical results have minor deviations  from the analytical, but the difference is explained by the non-linearity of dispersion. For example, from the Eq. (\ref{S_2^2Expansion}) we expect, that in the case of strictly linear dispersion, the term $q_z^2 b_z^2 $ should have zero 'bulk' contribution and small positive contribution from the Fermi surface. However, since on a lattice, the dispersion (\ref{Type_2_DM_Lattice}) is non-linear, it has indeed a large negative 'bulk' contribution responsible for the shape of the curves on the Fig.  \ref{Bz_plot}. Similarly, the wavelength of spin density waves 
is determined by scales of the band. In realistic materials, we expect non-linear corrections to play weaker role than in our simulations, since their scale is much larger than the chemical potential, but our main conclusion is that non-linear corrections to the band structure may change magnetic ground states in numerous ways, even though they are still expected to be spatially modulated.

Now, let us repeat the calculations in the case of type \Romannum{1} Dirac semimetal. In a similar way, we compute the effective action (\ref{TrGMGM_gen_expr})
for the model (\ref{HamiltonianTypeOne}) placed on a lattice. We  
choose the functions $d$ as:
\begin{eqnarray}
d_1 &=& \sin k_x,
\nonumber\\
d_2 &=& \sin k_y,
\nonumber\\
d_3 &=& m_0 - m_1 (1 - \cos k_z).
\label{Type_1_DM_Lattice}
\end{eqnarray}

\begin{figure}
\centering
	\begin{subfigure}[t]{0.5\textwidth}
		\includegraphics[width=9cm,angle=0]{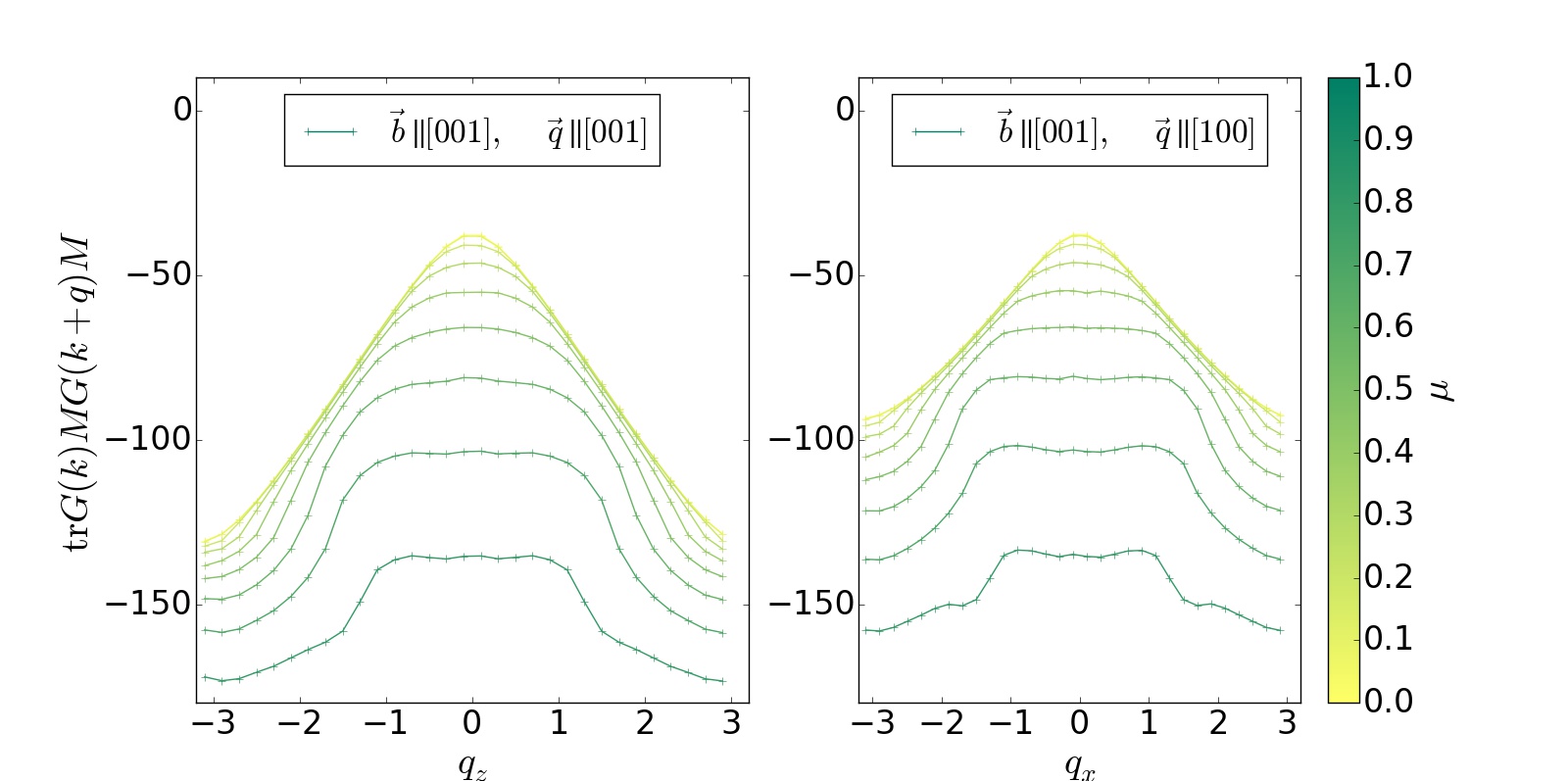}
		\subcaption{$\vec{b} \parallel \vec{z}$}	
		\label{Bz_plot_1}
	\end{subfigure}
	\begin{subfigure}[t]{0.5\textwidth}
		\includegraphics[width=9cm,angle=0]{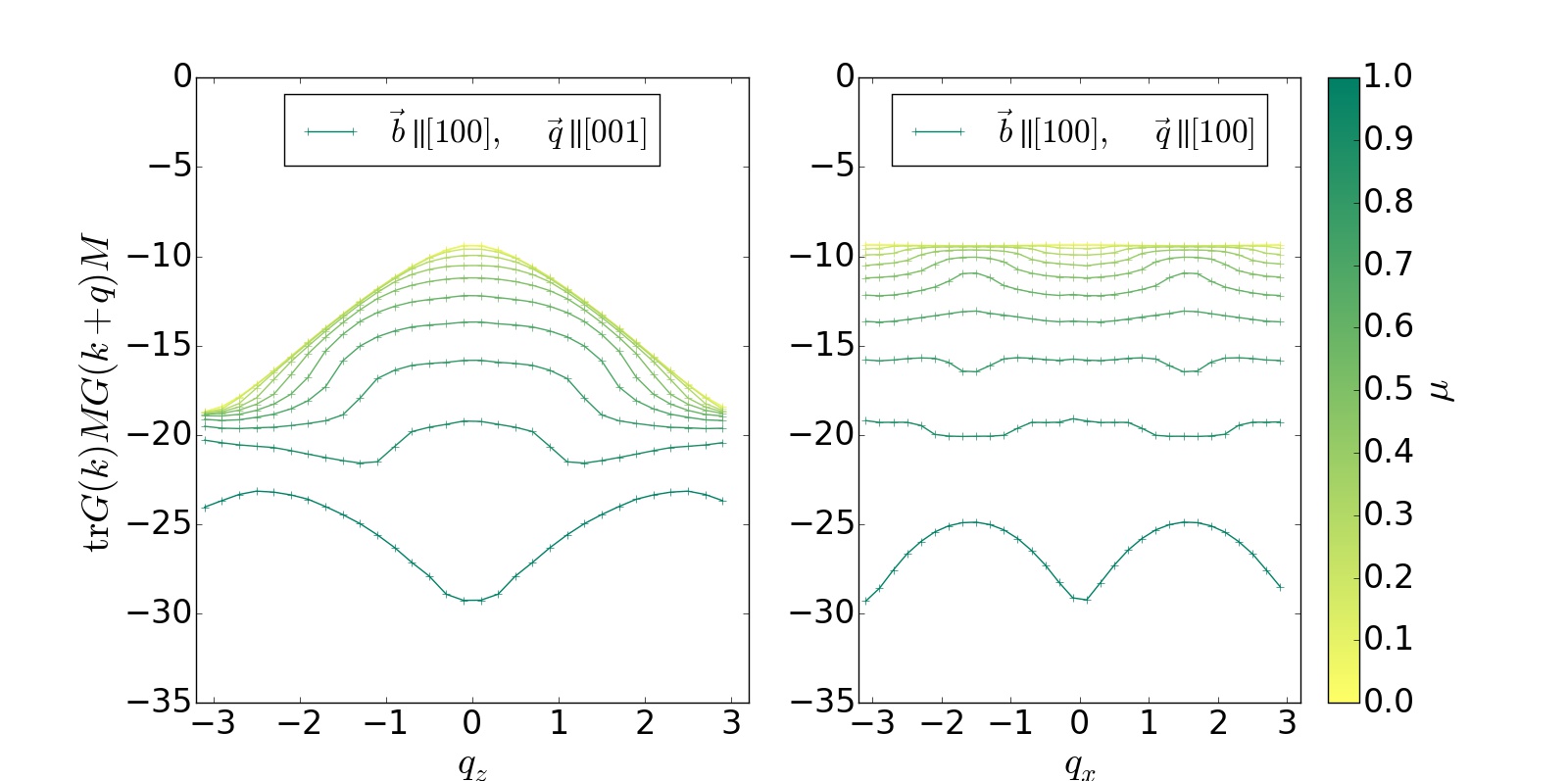}
		\subcaption{$\vec{b} \parallel \vec{x}$}	
		\label{Bx_plot_1}
	\end{subfigure}
\caption{
	Numerically computed effective action for the model (\ref{Type_1_DM_Lattice}) as functions of $q_z$, $q_x$. The parameters of the model are: $m_0 = 1.0$, $m_1 =1.0$, $g_s = g_p = 1.0$. The magnetic field is: (\ref{Bz_plot_1}) $\vec b = (0, 0, 0.5)$, and  (\ref{Bx_plot_1}) $\vec b = (0.5, 0, 0)$. One can see, that $b_z$ has minima at either $p_x=\pi$ or $p_z=\pi$, which suggests that the magnetic ground state is spatially modulated. 
 Also, one can see (\ref{Bz_plot_1}) that at large $\mu$, the local minimum gets displaced away from the boundary of the Brillouin zone. Similarly, one can see that $b_x$ forms an antiferromagnetic configuation modulated in $x$ direction, but with increasing $\mu$, it changes into spin density wave. 
}
\label{Fig:TypeOne}
\end{figure}

We present our findings on the Fig. \ref{Bx_plot_1}. One can see that in the case of magnetization in $z$ direction, the effective action has local minima either at $X$, or at $Z$ point of the Brillouin zone (i.e. at $q= (\pi,0,0)$ or $q= (0,0,\pi)$ ), which suggest that $b_z$ can be spatially modulated 
in both $x$  and $z$ directions. On the other hand, $b_x$ is modulated in $z$ direction: at small $\mu$, it forms an antiferromagnet, but as $\mu$ increases, a transition to spin density wave phase occurs. Eventually, at large $\mu$, a phase transition occurs: spatial modulation in $z$ direction disappears, and the system becomes spatially modulated in the $x$ direction.

\subsection{Magnetic ground states at finite temperatures}
\label{Lattice_finite_temperature}

Once we found the magnetic ground states at zero temperature, we can try to study their evolution, once temperature becomes finite. Specifically, we would like to compute numerically the effective action (\ref{TrGMGM_gen_expr}) at finite temperatures. Since $n_F$ and its derivative 
significantly deviate from constant only in a narrow range of parameters, straightforward tetrahedral integration is challenging. For this reason, to obtain the effective action at finite temperature, we first compute it for various  chemical potentials at zero temperature, and then obtain the answer at finite temperatures by applying a discretized version of the following relation \cite{PhysRevB.49.16223}:
\begin{eqnarray}
S (T, \mu) = - \int d \xi \frac{\partial n_F(\xi -\mu)}{\partial \xi}  S (0, \xi).
\label{Finite_T_vs_Zero_T}
\end{eqnarray}

In addition, we have to account explicitly for the change of chemical potential with temperature, which we do in the following way. First, we numerically compute particle number at zero temperature as a function of chemical potential: $N = \int \frac{d^3 k}{(2\pi)^3} \theta(\mu - k)$. Then,  by using  Eq. (\ref{Finite_T_vs_Zero_T}), we obtain particle number  $N(\mu, T)$ at various finite temperatures and chemical potentials and use it to find the dependency $\mu(T)$. Namely, for a given chemical potential at zero temperature $E_F$, we take number of particles $N(E_F, T=0)$, then, for given nonzero $T$, find two closest values of $N(T)$ and their corresponding $\mu$  and finally obtain the answer for $\mu(T)$ through their linear interpolation. 
In total, we find $\mu(T)$ for given $E_F$ and then use it to compute the effective action through  Eq. (\ref{Finite_T_vs_Zero_T}). 



\begin{figure}
\centering
	\includegraphics[width=9cm,angle=0]{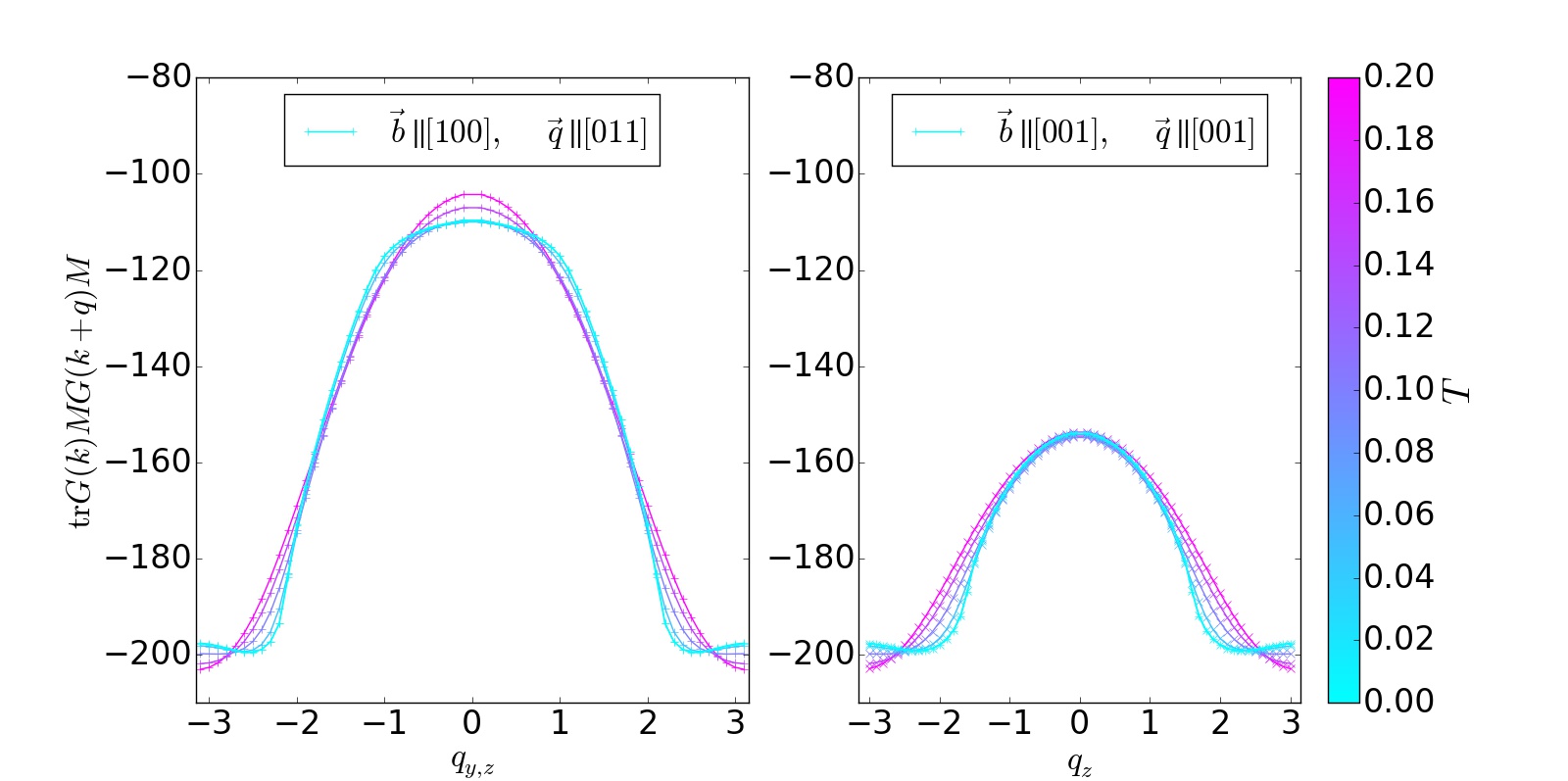}
\caption{
	Numerically computed effective action for the model (\ref{Type_2_DM_Lattice}) at various temperatures. The magnetic field and momentum have are directed: (left) $b= (b_x, 0, 0)$, $q = (0, q, q)$; (right) $b = (0, 0, b_z)$, $q = (0, 0, q_z)$. We choose $\mu = 0.7$, and the other parameters are the same as on the Fig. \ref{Fig:TypeTwo}.
One can see that, at zero $T$, the effective action has minumum away from the boundary of the Brillouin zone, but as $T$ increases, the minimum shifts towards it. Thus the system undergoes a phase transition from antiferromagnetic to spin density wave phase.}
\label{Fig:TypeTwoTemperatures}
\end{figure}

In the case of type \Romannum{2} Dirac semimetal, we present our findings on the Fig. \ref{Fig:TypeTwoTemperatures}. At $q=0$ we expect that the effective action will be temperature-independent for $b \parallel z$, but will increase e.g. for $b \parallel x$. Our numerical plots are consistent with these predictions at small $T$, but as $T$ becomes large, the numerical plots start behaving differently because of nonlinearities in the dispersion. More interestingly, our numerical findings confirm that  wavevector of the magnetization changes with temperature, and even more, as temperature grows, the system may undergo a phase transition from spin density wave to antiferromagnetic phase. 

\begin{figure}
\centering
	\includegraphics[width=9cm,angle=0]{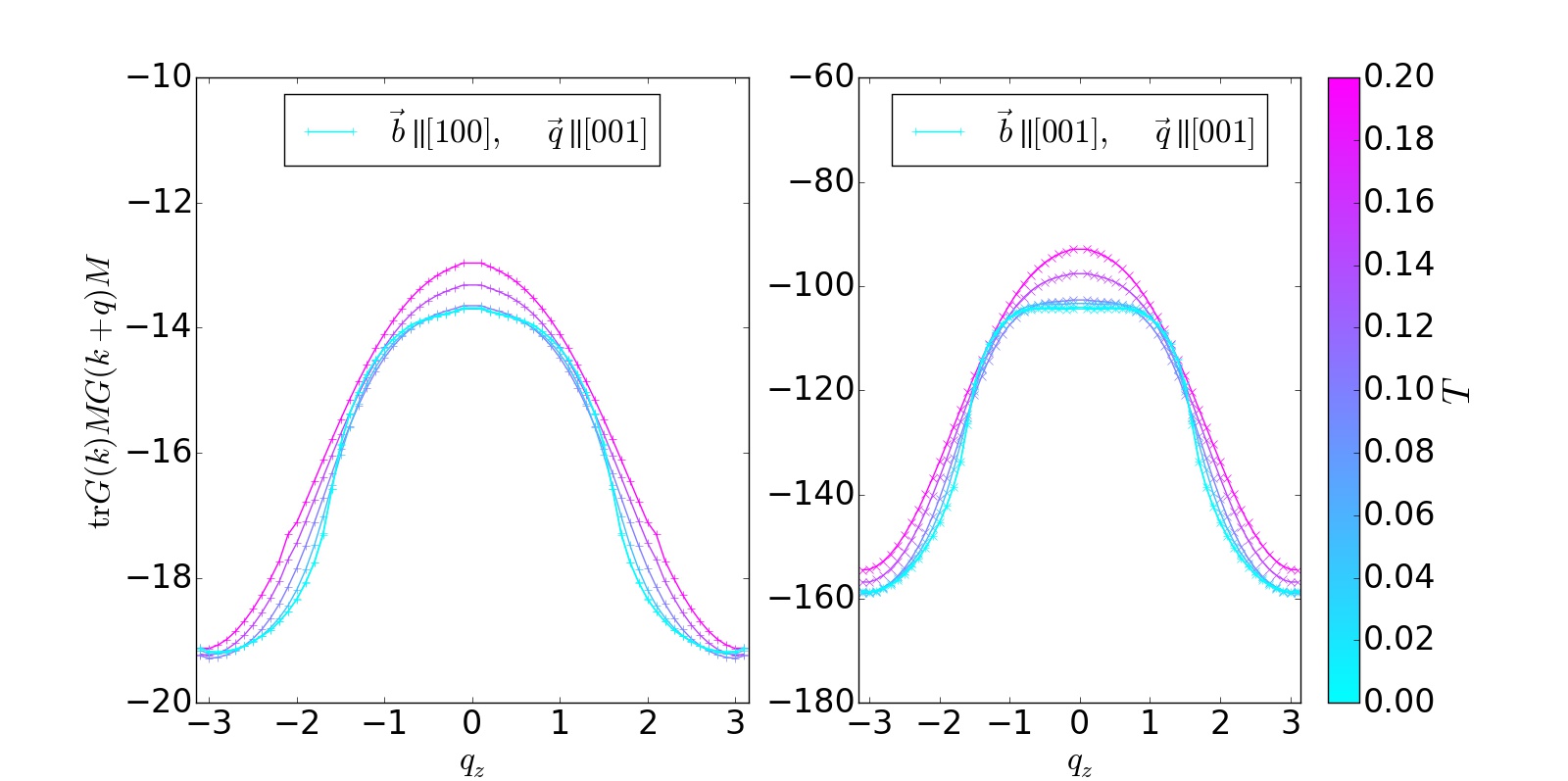}
\caption{
	Numerically computed effective action for the model (\ref{Type_1_DM_Lattice}) at various temperatures. We choose: (left) $b = (b_x, 0, 0)$, $q = (0, 0, q_z)$; and (right) $b = (0, 0, b_z)$, $q = (0, 0, q_z)$. Here $\mu=0.7$ and the other parameters are the same as on the Fig. \ref{Fig:TypeOne}.
}
\label{Fig:TypeOneTemperatures}
\end{figure}

The results for  type \Romannum{1} Dirac semimetal are qualitatively similar to type \Romannum{2} case and  are shown on the Fig. \ref{Fig:TypeOneTemperatures}. As expected, the magnitude of the effective action decreases with temperature, and its shape evolves. Specifically, its minimum may shift, so that the system undergoes a transition from spin density wave to antiferromagnetic phase. 

Thus, we have demonstrated that Dirac electrons may have a large variety of possible magnetic phases. Most commonly, these phases are antiferromagnetic or spin density wave, and there are possible transitions between them, as chemial potential or temperature varies. 

\newpage

\bibliographystyle{apsrev4-1}
\bibliography{Bib_Dirac_Magnetism,Bib_Dirac}

\end{document}